# Labyrinthine acoustic metamaterials with space-coiling channels for low-frequency sound control

**A.O. Krushynska[1,*], F. Bosia[1], N.M. Pugno[2,†]**

[1]Department of Physics and Nanostructured Interfaces and Surfaces, University of Torino, Via Pietro Giuria 1, 10125 Torino, Italy

[2]Laboratory of Bio-Inspired and Graphene Nanomechanics, Department of Civil, Environmental and Mechanical Engineering, University of Trento, Via Mesiano 77, 38123 Trento, Italy

17 December 2017

*akrushynska@gmail.com

[†]nicola.pugno@unitn.it Also at:

(i) School of Engineering and Materials Science, Queen Mary University of London, Mile End Road, London E1 4NS, United Kingdom;

(ii) Ket Labs, Edoardo Amaldi Foundation, Italian Space Agency, Via del Politecnico snc, 00133 Rome, Italy

**Abstract**

We numerically analyze the performance of labyrinthine acoustic metamaterials with internal channels folded along a Wunderlich space-filling curve to control low-frequency sound in air. In contrast to previous studies, we perform direct modeling of wave propagation through folded channels, not introducing effective theory assumptions. As a result, we reveal that metastructures with channels, which allow wave propagation in the opposite direction to incident waves, have different dynamics as compared to those for straight slits of equivalent length. The differences are attributed to activated tortuosity effects and result in 100% wave reflection at band gap frequencies. This total reflection phenomenon is found to be insensitive to thermo-viscous dissipation in air. For labyrinthine channels generated by iteration levels, one can achieve broadband total sound reflection by using a metamaterial monolayer and efficiently control the amount of absorbed wave energy by tuning the channel width. Thus, the work contributes to a better understanding of labyrinthine metamaterials with potential applications for reflection and filtering of low-frequency airborne sound.

Keywords: low-frequency waves, labyrinthine acoustic metamaterial, space-coiling curve, Fabry-Perot resonance, hierarchical organization, perfect reflector, tortuous porous material.

## 1. Introduction

Acoustic metamaterials are composites with an engineered structure governing remarkable functionalities, e.g. acoustic cloaking, transformation acoustics, and subwavelength-resolution imagining [1, 2]. Apart from unusual effective properties, the metamaterials offer various possibilities to control propagation of sound or elastic waves at deep sub-wavelength scales [3, 4, 5]. This can be achieved by incorporating heavy resonators [3], Helmholtz resonators [6, 7], tensioned membranes [8, 9], or sub-wavelength perforations or slits [10, 11, 12, 13] in a material structure. A class of acoustic metamaterials with internal slits is also known as "labyrinthine". They have recently attracted considerable attention due to their abilities to exhibit an exceptionally high refractive index and efficiently reflect sound waves, while preserving light weight and compact dimensions [13, 12, 14].

Labyrinthine metamaterials enable to slow down the effective speed of acoustic waves due to path elongation by means of folded narrow channels [15, 13]. Their high efficiency in manipulating low-frequency sound has been experimentally demonstrated for various channel geometries. For example, Xie et al. [16] have shown the appearance of a negative effective refractive index at broadband frequencies for labyrinthine metastructures with zig-zag-type channels. For the same configuration, Liang et al [15] have demonstrated extraordinary dispersion, including negative refraction and conical dispersion for low-frequency airborne sound. Frenzel et al. have used the zig-zag channels to achieve broadband sound attenuation by means of three-dimensional labyrinthine metastructures [17, 18]. The issue of poor impedance matching for labyrinthine metamaterials has been addressed by exploiting tapered and spiral channels [19] and hierarchically structured walls [20]. Cheng et al. have proven almost perfect reflection of low-frequency sound by sparsely arranged unit cells with circular-shaped channels that can induce artificial subwavelength Mie resonances [12]. In our previous work, we have proposed a simple modification to the latter design (by adding a square frame) to achieve a wider tunability [14]. Moleron et al. have emphasized the importance of thermo-viscous effects on the performance of labyrinthine structures with sub-wavelength slits [21].

Most of the studies analyze labyrinthine metamaterials with curved channels by replacing a real system with a simplified one, when dynamics of folded channels is described by that of straight slits of an effective length, which equals to the shortest path taken by a wave within the structure [13, 21, 17, 15, 20]. This approach provides reliable results for channels, in which the direction of wave propagation does not deviate much from that that for incident waves. Therefore, it appears that the channel tortuosity plays no role. Possible effects of the path tortuosity, when a wave is allowed to propagate in the opposite direction relative to that of the incident field, remain to be investigated. A limited number of papers have analyzed labyrinthine metamaterials of this type. In [19], Xie et.al have investigated metastructures with spiral channels to introduce tunability of effective structural parameters, such as refractive index and impedance. Song et. al. have considered hierarchically organized walls to achieve a broadband wave absorption [20]. These works are mainly focused on the experimental validation of the mentioned features, and lack a theoretical analysis of wave behavior in a tortuous channel.

The goal of this work is to numerically investigate dispersion and propagation properties of airborne sound in labyrinthine metamaterials with channels that allow a change in the direction of wave propagation, and compare their performance with that of the corresponding straight slits. For this purpose, we design sub-wavelength paths in metamaterial unit cells along a hierarchically-organized curve. In particular, we consider a space-filling curve due to its self-similar organization, a simple algorithm to derive length elongation, and the inherent property to fill in an occupied area. We provide a complete theoretical analysis of the wave dispersion in the designed metamaterials complemented by the study of acoustic transmission, reflection, and absorption for a monoslab in the absence or presence of thermo-viscous losses. Our results demonstrate that, when a wave inside a narrow channel is allowed to propagate in the opposite direction with respect to the incident wave front, the wave dynamics is not equivalent to that in a straight slit of an effective length. The peculiar channel tortuosity allows to open wide sub-wavelength band gaps. At band gap frequencies, total broadband wave reflection occurs that is not influenced by the presence of losses in air. Therefore, the proposed labyrinthine metamaterials provide great potential as efficient reflectors for low-frequency airborne sound. Moreover, to facilitate practical exploitation of these metamaterials, we propose to assemble reconfigurable structures from constant thickness thin panels (sheets), which provide a cheap alternative to an additive manufacturing approach.

## 2. Space-filling curves

As mentioned above, the wave path is elongated by exploiting the hierarchical structure of space-filling curves [22]. First space-filling curves were discovered by Peano [23] (later named after him), and since then many other curves were proposed [24]. An attractive property of these curves is that they go through every point of a bounding domain for an unlimited number of iterations. After initially being studied as a curiosity, nowadays space-filling curves are widely applied, e.g. for indexing of multi-dimensional data [25], transactions and disk scheduling in advanced databases [26], building routing systems [27], etc.

Among various space-filling curves, we have chosen the Wunderlich two-dimensional curve filling a square [22], which is constructed as follows. At the $1^{st}$ iteration level, one draws an "S"- shaped curve starting at the bottom-left corner of a bounding square and ending at the top-right corner. At the $n^{th}$ ($n \geq 2$) iteration level, 3 copies of the $(n$ -$1)^{th}$-level curve are arranged along each side of a square with every copy being rotated by 90° relative to the previous one. The curves are joined into an N-shaped route starting from the up-direction for the left column, then down for the middle column, and finally again up for the right column. At every iteration level $N$, the length of the Wunderlich curve is $(3^N - 1/3^N)$, while that of e.g. Hilbert's curves is $(2^N - 1/2^N)$ [22]. Faster length elongation enables more compact channel folding in a labyrinthine structure (and thus, increases the tortuosity effect, as will be shown later) that justifies the choice of the Wunderlich curve for this study.

## 3. Models and analysis methods

Figure 1 presents square labyrinths with an internal channel shaped along the Wunderlich curve of the three iteration levels, which are used for constructing "unit cell 1" (UC1), "unit cell 2" (UC2), and "unit cell 3" (UC3), respectively. The structural material is aluminum with mass density $\rho_{Al} = 2700$ kg/m³ and speed of sound $c_{Al} = 5042$ m/s. The thickness of bounding walls is fixed for all the unit cells and equals $d$=0.5mm.

The channel width is $w$, and the size of a square domain occupied by a single labyrinth is $a = 3^N \cdot (w + d) + d$, where $N$ is the iteration level. We preserve an interconnecting cavity of width $w$ between adjacent labyrinths. Thus, the metamaterial unit cell size is $a_{uc} = a + w$ (see Fig. 1a for notations).

We analyze plane waves propagating in the plane of a unit cell cross-section. The metamaterial geometry is assumed to be constant in the out-of-plane direction without a possibility to excite a momentum in this direction. Hence, the pressure field is always constant in the out-of-plane direction, and the wave dynamics can be analyzed by considering a two-dimensional (2D) geometry. The validity of this assumption is confirmed by a good agreement with the results of three-dimensional (3D) simulations given further in the Section 4.

First, we analyze sound wave dispersion in the labyrinthine metamaterials that are infinite in both in-plane directions. By neglecting any losses in air, small-amplitude variations of harmonic pressure $p(\boldsymbol{x},t) = p(\boldsymbol{x})e^{i\omega t}$ (with angular frequency $\omega = 2\pi f$, where $f$ the frequency in Hz) are governed by the homogeneous Helmholtz equation:

$$\nabla \cdot \left(-\frac{1}{\rho_0}\nabla p\right) - \frac{\omega^2 p}{\rho_0 c_0^2} = 0 \qquad (1)$$

with air density $\rho_0 = 1.225$ kg/m$^3$ and speed of sound $c_0 = 343$ m/s at a temperature of $T = 20°\text{C}$. Since the characteristic acoustic impedance of aluminum is around 4 orders of magnitude larger than that of air, we assume zero displacements for the structural walls and apply sound-hard boundary conditions at air-structure interfaces. The pressure distribution at opposite unit cell boundaries is constrained by the Floquet-Bloch periodic conditions:

$$p(\boldsymbol{x} + \boldsymbol{a}) = p(\boldsymbol{x})e^{i\boldsymbol{k}\cdot\boldsymbol{a}} \qquad (2)$$

with $\boldsymbol{a} = (a_{uc}, a_{uc}, 0)$ and wave vector $\boldsymbol{k} = (k_x, k_y, 0)$. More details about the dispersion analysis can be found in [14].

Next, we evaluate homogeneous wave propagation through a metamaterial monolayer. Sketch of the model is presented in Fig. 2. Plane wave radiation occurs at the left domain boundary at distance of $10a_{uc}$ from the slab. At the right boundary, a perfectly matched layer of width $2a_{uc}$ is added to eliminate unwanted wave reflection. At the bottom and top boundaries, the Floquet-Bloch periodic boundary conditions (2) enable to artificially extend the air domain in the vertical direction. The reflection $R = |p_r/p_i|^2$, transmission $T = |p_t/p_i|^2$, and absorption $A = 1 - R - T$ coefficients are evaluated by averaging incident $p_i$, reflected $p_r$, and transmitted $p_t$ pressure fields along the lines located at distance $a_{uc}$ from the structure.

In order to analyze how the tortuosity of a labyrinthine channel influences sound wave characteristics, we compare the evaluated $T$ and $A$ values for the metastructures with those for straight slits of width $w$ between solid blocks of length $L$=$L_{eff}$ or $L$=$a_{uc}$ distributed at distances $a$ along the vertical direction. In the case of $L$=$a_{uc}$, the blocks are of the same size as labyrinthine structures, but do not contain internal channels. The effective channel length $L_{eff}$ is approximately equal to the shortest wave path from the input to the output through a labyrinthine channel (as shown e.g. by light-blue lines in Fig. 1b).

If a channel width is small compared to the wavelength of a propagating wave, thermal and viscous boundary layers near walls cause loss effects (*lossy* air). The thickness of these layers decreases with increasing frequency. The thickness of thermal boundary layer $\delta_{th}$ is evaluated as follows:

$$\delta_{th} = \sqrt{\frac{k}{\pi f \rho_0 C_p}}, \tag{3}$$

where $k = 25.8$ mW/(m·K) is the thermal conductivity, and $C_p = 1.005$ kJ/(m³·K) is the heat capacity at constant pressure. The thickness of the viscous boundary layer $\delta_{vis}$ is

$$\delta_{vis} = \sqrt{\frac{\mu}{\pi f \rho_0}}, \tag{4}$$

with dynamic viscosity $\mu = 1.814e\text{-}5$ Pa·s. The graphical representation of Eqs. (3)-(4) is given in Fig. 3. At 20°C and 1 atm, the viscous and thermal boundary layers are of thickness 0.22mm and 0.26mm at 100 Hz, respectively.

As the designed labyrinthine channels are relatively easy to model, we directly include thermal conduction and viscous attenuation into the governing equations. Thus, the linearized system consists of a Navier-Stokes equation, a continuity equation, and an energy equation, which are given in [28]. This system is solved for acoustic pressure variations $p$, the fluid velocity variations $\boldsymbol{u}$, and the acoustic temperature variations $T$. The variations describe small harmonic oscillations around a steady state. The mentioned equations are implemented in Thermoacoustic interface of Comsol Multiphysics [29].

The dispersion and transmission analyses are performed as eigenvalue and frequency-domain finite-element simula-tions. The described acoustic domains are discretized with the maximum element size of $\lambda_{min}/12$, where $\lambda_{min} = c_0/f_{max}$, and $f_{max}$ is the maximum considered frequency. Such a mesh resolves the smallest wavelength of the study with 12 elements. To properly capture the wave field variations within the viscous and thermal boundary layers, we implemented a frequency-varying mesh with 3-5 boundary layers along the thickness of the viscous layer.

## 4. Results and discussion

We consider the designed labyrinthine metamaterials of two dimensions. In the first case, defined as a "fixed channel" case, we imply a constant channel width, $w = const$, at each iteration step. Thereby we aim at evaluating effects of tortuosity on sound propagation in elongating paths. For $w$=4 mm, the metamaterial unit cell sizes are $a_{uc}$ =18 mm for UC1, 45 mm for UC2, and 126 mm for UC3. For another case, called as "fixed unit cell" case, we assume a fixed unit cell size, $a_{uc} = const$, with the channel width becoming smaller at each iteration. In particular, we fix $a_{uc} = 14$ mm that corresponds to the channel width $w = 3$ mm for UC1 and 0.9mm for UC2. For UC3, the internal channel is disappears for the specified wall thickness $d$=0.5 mm. For the chosen value of $a_{uc}$, the channel width in the "fixed unit cell" case is smaller than that in the "fixed channel" case at the same iteration level. Thus, by comparing wave propagation for these two cases, we can evaluate how different amount of thermo-viscous losses influences the wave dynamics in labyrinthine channels of the same structure.

In the both described cases, an internal labyrinthine channel is shaped along the Wunderlich curve. However, its length is scaled differently than that of the fractal curve due to deviations in construction approaches. Specifically, the algorithm of the Wunderlich curve construction assumes that the curve is a compressing mapping from a low-dimensional space into a 2D domain, the area of which is the same at each iteration level [22]. For our unit cells, we assume the constant wall thickness that incurs variations in the channel length relative to that of the Wunderlich curve. Hence, in the "fixed channel" case, when the area of a bounding square increases at each iteration step (in contrast to the construction approach of the Wunderlich curve), the channel length is elongated by a factor of $3^N$ relative to $a$. In the "fixed unit cell" case, the channel length increases as $3^N a - 1$.

### 4.1 "Fixed-channel" case

Figure 4 shows evaluated dispersion relations for homogeneous waves in UC1, UC2, and UC3 propagating along $\Gamma$X direction in the reciprocal $k$-space. The horizontal axis indicates normalized wavenumber $k^* = a_{uc}k$, and the vertical axes depict frequencies $f$ in kHz and normalized frequencies $f^* = f a_{uc}/c_0$. Note different frequency ranges for each unit cell. The analyzed frequencies are limited to a sub-wavelength range, i.e. up to $f a_{uc}/c_0 = 0.5$. For UC1, we consider modes forming the lowest band gap and extending up to 9 kHz. For the UC2 and UC3, the frequency range includes the first 4 band gaps, and thus, it is limited to 4 kHz and 500 Hz, respectively.

The dash-dot lines represent phase velocities of sound waves in lossless air for the lowest fundamental mode within a unit cell (green curve) and in homogeneous air, when a unit cell is removed (red curve). As can be expected, the velocity is reduced when a wave propagates through a labyrinthine channel. The reduction factor is 1.63 (UC1), 2.91 (UC2), and 5.28 (UC3) compared to homogeneous air.

The dispersion relations in Fig. 4 are characterized by several frequency band gaps in the sub-wavelength region. Hence, the designed labyrinthine metamaterials can control sound waves at sub-wavelength scales. As $N$ increases, the band gaps are shifted down to lower frequencies. The shifts are directly related to the path elongation. For example, the 1st band gap starting from $fa/c_0 = 0.21$ for UC1, is shifted to about 3 times lower frequency, $fa/c_0 = 0.069$, for UC2, since the channel length in UC2 is 3 times longer than that in UC1.

The band gaps bounds are formed by flat parts of dispersion bands that correspond to localized modes. The pressure distributions for these modes are given in the 1st and 3rd columns of Table 1 for the 1st band gap bounds and Table 2 for the 2nd and 3rd band gap bounds. Red and blue colors represent maximum and minimum values of pressure, while green color indicates near-zero pressure. Strong pressure localization is observed within the labyrinthine channels. It is easy to estimate that regardless of the iteration level, these localized modes correspond to Fabry-Perot resonances in a straight slit of width $w$ and length $L_{eff}$ [21, 13]:

$$f_l^{FP} = lc_0/2L_{eff}, \qquad (5)$$

where $l$ is a positive integer. In the "fixed channel" case, $L_{eff}$ equals $2.305d_{uc}$ for UC1; $L_{eff} = 5.667d_{uc}$ for UC2, and $L_{eff} = 16.642d_{uc}$ for UC3 with $d_{uc} = a_{uc}\sqrt{2}$. Note that odd $l$ values correspond to the lower band gap bounds, while even $l$ values allow approximating the upper band gap bounds in Fig. 4.

The fact that multiple Fabry-Perot resonances form the band gap bounds explains a similar structure of the dispersion bands at various frequencies in Fig.4, which have close values of phase and group velocities.

The pressure distributions given in Tables 1-2 also resemble those of artificial monopole, dipole and multipole resonances from Ref. [12]. For example, the patterns at lower bound of the 1st band gap (the 1st column in Table 1) is similar to a monopole, in which the pressure is concentrated in the central part of a channel, equally radiating along two propagation directions [12, 14]. Thus, the monopole and multipole resonances in the considered folded channels originate from the tortuosity effect of the Fabry-Perot resonances.

Since an effective dynamic bulk modulus (not evaluated in this study) is typically negative at limited frequencies above the monopole resonance, one can expect a high wave reflectance at these frequencies [12]. This behavior has been experimentally observed in [12] for circular-shaped folded channels. The wave transmission and absorption coefficients for our metastructures are discussed below in this section.

Apart from the Fabry-Perot resonances, wave dispersion in the designed labyrinthine metamaterials is also characterized by the presence of bands within the band gap frequencies. These bands are found within each band gap for every analyzed unit cell (see Fig. 4). The pressure distributions for these modes (the 2nd column in Tables 1-2) resemble those for the dipole and its higher harmonics (compare to 3$^{rd}$ column of Tables 1-2), but it is not localized inside a channel. Hence, these modes do not represent standing localized waves, rather they are propagating waves with very small (and often negative) group velocities. They may be analogous to slow modes inside phononic band gaps for elastic waves [30, 31]. The mechanism of the slow mode excitation in acoustics and their dynamics will be investigated in more detail in future work. Here, we consider these modes to be included in a single band gap (instead of separating a band gap into two parts), since we have not detected the presence of these modes in the frequency-domain simulations (for lossless and lossy air), even for a very fine frequency step (see Figs. 5-6).

Frequency-domain simulation results are given in Figs. 5-6 in terms of transmission and absorption coefficients for lossless and lossy air. (Reflection coefficient can be directly derived from these data, and thus is not shown here.) We analyze waves propagating through a monolayer composed of the labyrinthine unit cells (Figs. 5a, 6a, 6c) and periodic straight slits of length $L_{eff}$ (Fig. 5b, 6b, 6d) or $a_{uc}$ (Fig. 5c). Note that at certain frequencies for lossy air, the transmission and absorption coefficients appear to be mesh-dependent, and hence are not shown here as unreliable.

When losses in air are neglected, incoming waves are either transmitted or reflected for all the considered geometries, and thus, the absorption coefficient is zero (not shown in the graphs). Total transmission is achieved at frequencies of the Fabry-Perot resonances given by Eq. (5). As can be seen, this effect is independent of the channel tortuosity and occurs in folded labyrinthine channels of any iteration level at almost the same frequencies as for straight slits. For the slit of length $a_{uc}$, the fundamental Fabry-Perot resonance appears to be at higher frequencies than the analyzed frequency range. Thus, straight slits of length $a_{uc}$ will be not considered further.

When thermo-viscous losses are included, the transmission peaks decrease in magnitude and are shifted to lower frequencies compared to the lossless air. The latter occurs due to the reduction of the propagation velocity in dissipative air and is confirmed by experimental measurements in [21].

The striking difference in wave propagation through the unfolded (straight) and hierarchically-structured channels occurs between the frequencies of Fabry-Perot resonances. In case of the straight slits, the main part of incoming waves is reflected, while about 15-20% of the wave energy is transmitted through a slit. For the labyrinthine metamaterials, the same behavior is observed in the propagating frequency range, while within the band gaps total wave reflection occurs with zero transmission coefficient. As mentioned above, at the lower band gap bound, the fundamental Fabry-Perot resonance corresponds to the monopole, and thus, total reflectance is justified by a negative value of effective bulk modulus within the band gap. Experimental data for circular-shaped folded channels [12] show about 84% insertion loss that is in good agreement with the transmission results for straight slits at frequencies between the Fabry-Perot resonances (see e.g. in Fig. 6b). In contrast to this, for our labyrinthine structures total zero transmission is achieved even if thermo-viscous losses are taken into account. We attribute this to the presence of a wave path that redirects a propagating wave in the folded channel to the opposite direction relative to incident waves, since all the other structural parameters (as compared to a straight slit) are the same. Therefore, the peculiar tortuosity of the designed channels significantly modifies the wave dynamics at band-gap frequencies, and these effects cannot be captured by a simplified consideration of equivalent straight slits.

While the total transmission at Fabry-Perot resonances is eliminated by the loss mechanisms in air [21], the revealed total reflection at band-gap frequencies is not affected by dissipation. As the iteration level increases, the band gaps, i.e. the total reflection frequencies, are shifted to lower frequencies and decrease in size (compare Figs. 5a, 6a, and 6c). However, the amount of transmitted energy at frequencies of propagating modes also decreases, which is not the case for the straight slits (compare e.g. Figs. 6c and 6d). Therefore, the incorporation of third and higher iteration levels for a "fixed channel" unit cell is beneficial for low-frequency sound control and allows to achieve broadband sound reflection.

To summarize, we can derive two key conclusions. First, wave propagation in the proposed labyrinthine metamaterials with hierarchically-structured channels differs from that through straight slits of the effective length. The physical mechanism causing this difference is the channel tortuosity, which allows a wave to propagate in the opposite direction relative to an incident pressure field. When deriving effective characteristics for metastructures with complex-shaped wave paths, the mentioned tortuosity effect must be taken into account. Second, the designed labyrinthine metamaterials can be used as broadband low-frequency sound reflectors of compact size, since 100% wave reflection is achieved by using a single unit cell.

The circular markers in Fig. 5a represent the transmission coefficient for a corresponding 3D domain obtained by extruding the 2D model (Fig.2) in the out-of-plane direction by a height of $4a_{uc}$. Excellent agreement between the 3D and 2D results justifies the introduced assumption of the two-dimensional character of the analyzed problem.

Finally, we note that the designed metamaterials can be compared with tortuous open-porous materials. The porosity level, evaluated as the ratio of the area of air inside a unit cell to the total area of a unit cell, is about 90% for UC1, 88 % for UC2, and 89 % for UC3, which is rather low as compared to porosity of typical foams slightly deviating from 100% [32]. The main difference between porous foams and the designed labyrinthine metamaterials is the physical mechanism of wave control. Porous materials attenuate waves due to inherent thermo-viscous losses with the absorption coefficient close to 1 for broad frequency ranges. In contrast to this, the proposed metastructures mainly reflect incident waves with absorption approaching 0.5 at single frequencies of Fabry-Perot resonances (see Figs. 6 a,c). In the next section, we estimate the metamaterial performance for an increased level of thermo-viscous losses.

## 4.2 "Fixed-unit-cell" case

In the "fixed unit cell" case, the unit cell size $a_{uc} = 14$ mm is fixed as for all the iteration levels. Dispersion relations of UC1 and UC2 are shown in Fig. 7 for homogeneous waves propagating along the $\Gamma X$ direction. The dimensional frequency ranges here are the same as those for the corresponding unit cells in the "fixed channel" case (see Figs. 4 a,b).

The structure of the dispersion relation in Fig. 7a is similar to that in Fig. 4a, except that the bands are shifted to higher frequencies. This occurs due to a shorter channel length. At first sight, more differences are found by comparing the dispersion relations for UC2 in Fig. 4b and Fig. 7b. While in Fig. 4b there are four band gaps, the relation in Fig. 7b is characterized by the presence of a single wide band gap. This happens because the unit cell area, $A^{(fix_{uc})} = 14^2$ mm$^2$, in the second case is about 3 times smaller than that for the "fixed channel case", $A^{(fix_w)} = 41^2$mm$^2$. As a result, the monopole, dipole and multipole resonances, as well as the related band gaps, are shifted to 3 times higher frequencies. However, in terms of non-dimensional frequencies, the band gap frequencies remain unchanged. The similarity of dispersion relations in Figs. 4 and 7 can be expected, since the metamaterial structure is preserved. In contrast to this, one should observe differences in transmission and absorption coefficients between these two cases due to the different amount of thermo-viscous losses in the channels of a various width.

Figure 8 shows the transmission and absorption coefficients for labyrinthine monoslabs of the "fixed unit cell" case and those for straight slits of the effective length $L_{eff}$ = 34.5 mm (UC1) and $L_{\text{eff}}$ = 107 mm (UC2). The key features found in the analysis of the "fixed channel" case are also observed in the present case, namely the wave propagation in the labyrinthine channels is not equivalent to that in straight slits due to the occurrence of 100% reflection within band gaps. The total reflection is again independent of losses in air. However, as the channel of UC2 in the "fixed unit cell" case is more than 4 times narrower relative to that in the "fixed channel" case, the influence of thermo-viscous losses becomes more pronounced. This can be primarily seen in larger absorption values at the Fabry-Perot resonant frequencies.

Therefore, wave attenuation within labyrinthine channels can be obviously increased by decreasing the channel width. The porosity of the metamaterial then also decreases. For UC2, the structural porosity is 64.7% for the "fixed unit cell" case versus 88% for the "fixed channel" case. Therefore, one can consider the wave absorption within the labyrinthine metamaterials as similar to that of tortuous foams by decreasing the structural external dimensions and porosity level.

## 5. Conclusions

In this work, we have theoretically analyzed the possibilities of labyrinthine acoustic metamaterials with sub-wavelength channels shaped along a space-filling curve to control airborne homogeneous sound. We have demonstrated that, if a folded channel allows wave propagation in an opposite direction compared to incident pressure, wave dynamics in the channel is not equivalent to that of a straight slit of an effective length. In particular, we have shown that Fabry-Perot resonances of the straight slit correspond to monopole, dipole and multipole resonances in folded channels and govern the generation of band gaps. Within the band gaps, total wave reflection occurs that is not influenced by the presence of dissipation in air. Moreover, by increasing the channel tortuosity and further elongating a wave path, one can achieve almost perfect reflection outside the band gaps. Although at higher iteration levels the designed metamaterials resemble a tortuous porous material, they mostly control waves due to interference effects, in contrast to thermo-viscous dissipation mechanism in porous foams. This results in a low wave attenuation within a metastructure for a sufficiently wide channel. The absorption level can be increased by decreasing the channel width and the structural weight.

This is the first time that a space-filling curve has been considered for designing and elongating wave paths in labyrinthine metamaterials. Therefore, further more in-depth analysis is required to study the influence of various geometric factors, e.g. number or angles of turns, as well as the metamaterial performance for inhomogeneous waves in complex-shaped folded channels. These studies will be performed in future works. The proposed structures show promise as broadband low-frequency sound reflectors that can be inexpensively assembled from thin sheets.

**Acknowledgements**

A.O.K. acknowledges Dr. Dmitry Krushinsky (University of Wageningen, the Netherlands) for the continuous support of this work and fruitful discussions. A.O.K. is supported by the funding from the European Union's 7[th] Framework programme for research and innovation under the Marie Skłodowska-Curie Grant Agreement No. 609402-2020 researchers: Train to Move (T2M). N.M.P. is supported by the European Commission under the Graphene FET Flagship (WP14 "Polymer

Composites" No. 604391) and FET Proactive "Neurofibres" grant No. 732344. FB is supported by "Neurofibres" grant No. 732344.

**Figures**

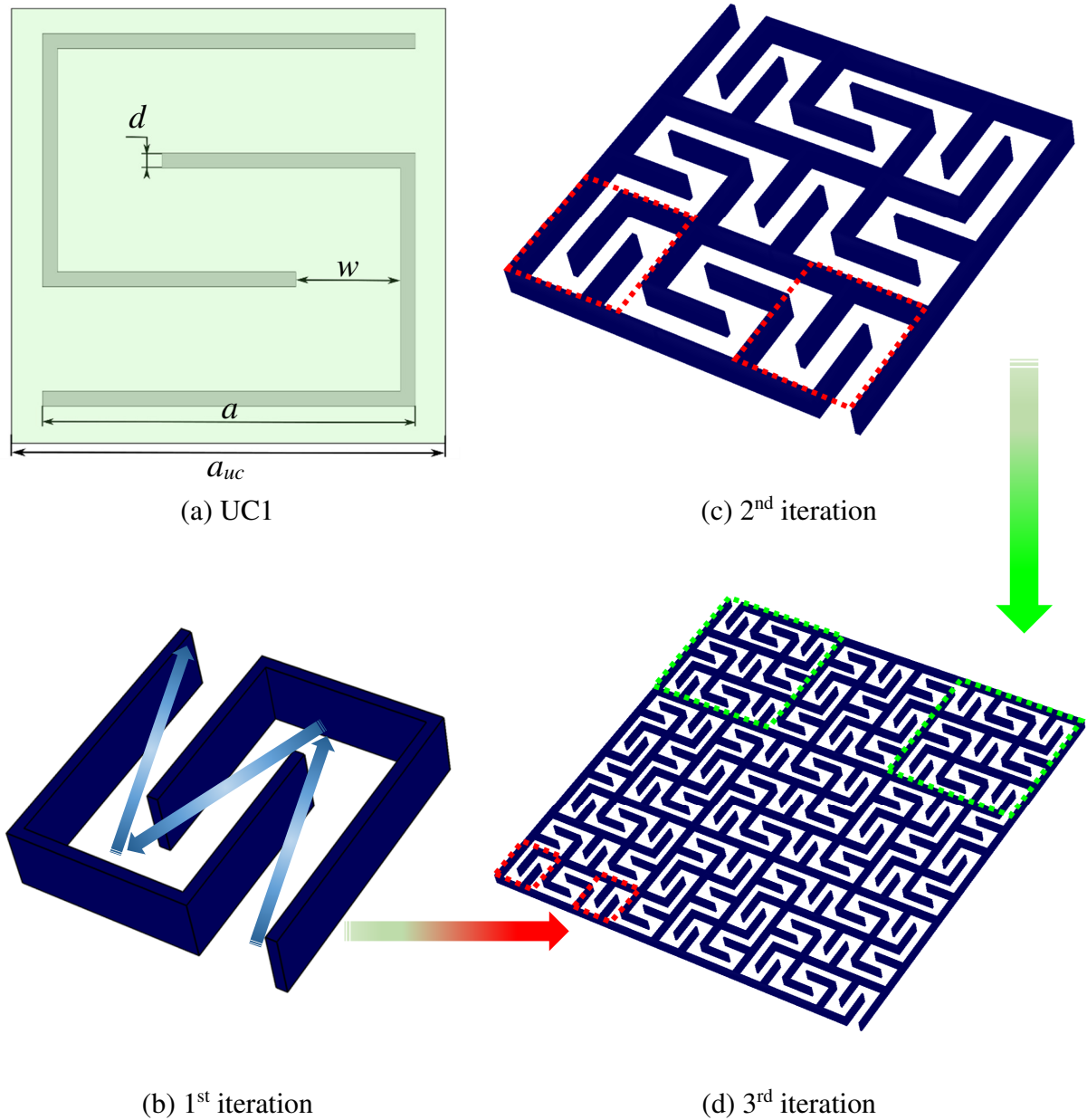

(a) UC1

(c) 2ⁿᵈ iteration

(b) 1ˢᵗ iteration

(d) 3ʳᵈ iteration

*Figure 1. (a) Unit cell of the 1ˢᵗ iteration level (UC1) with dimensions. (b-d) Labyrinths with air channels shaped according to the Wunderlich space-filling curve of the first three iteration levels incorporated into UC1, UC2, and UC3. Solid walls are indicated in blue. The shortest path taken by a wave within UC1 is shown by blue arrows in (b).*

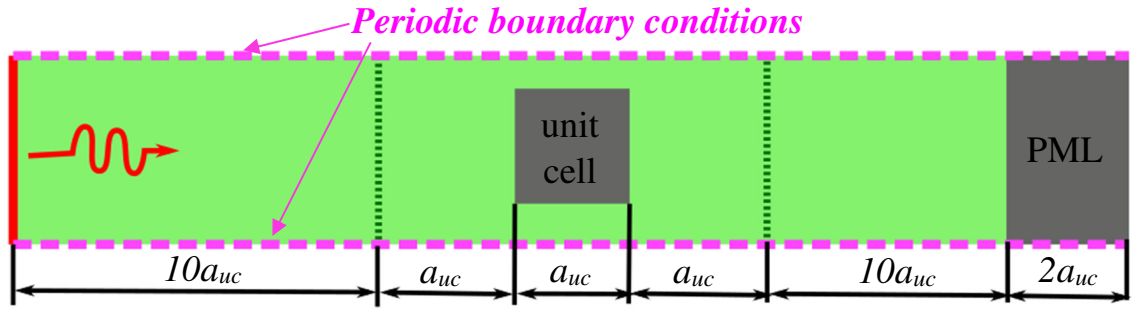

*Figure 2. Schematic of the frequency domain model. Green area corresponds to an air domain, green dashed lines indicate locations, at which reflection and transmission coefficients are evaluated. The plane wave radiation condition is applied along the bold red line.*

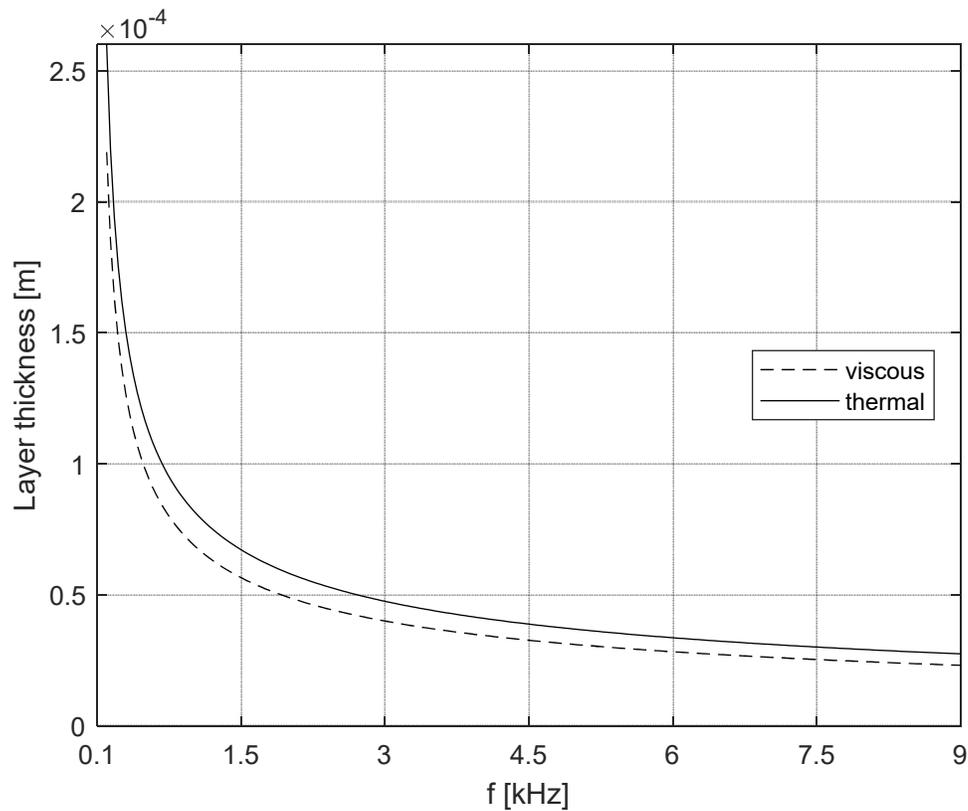

*Figure 3. Thickness of viscous $\delta_{vis}$ and thermal $\delta_{th}$ boundary layers according to relations (3) and (4).*

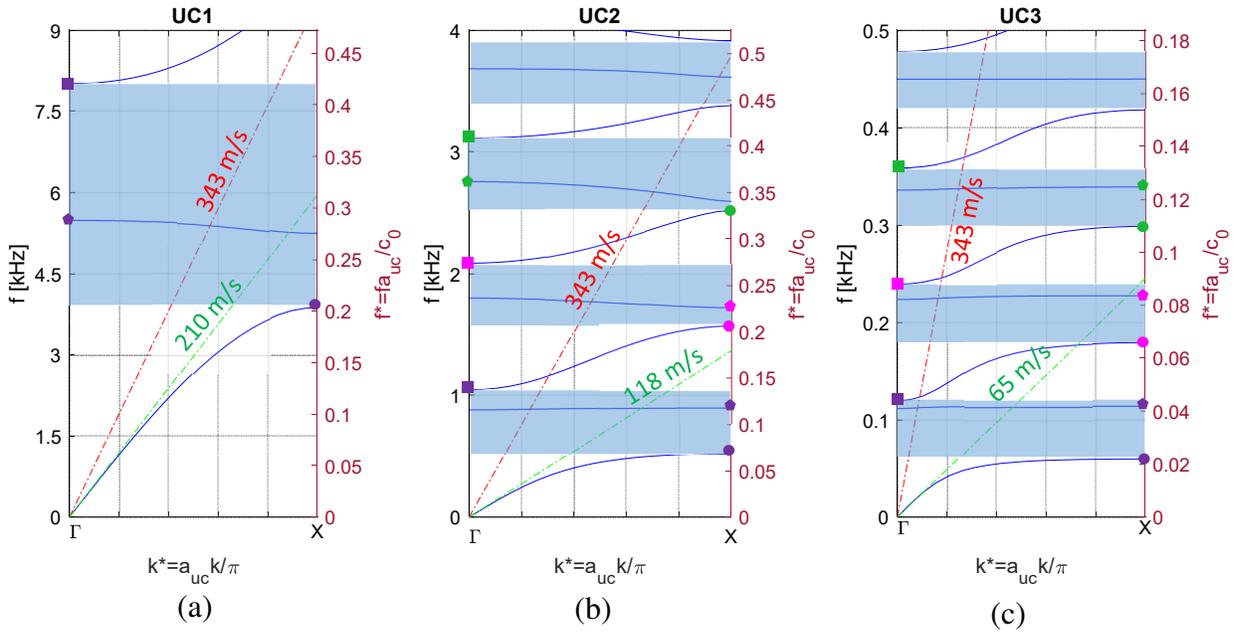

Figure 4. **"Fixed channel" case**: Dispersion relations for the labyrinthine unit cells of the 3 iteration levels with a fixed channel width, w=4 mm. Band gaps are shown by shaded rectangles. The slope of the green and red dash-dot lines indicates phase velocities of the fundamental mode within a unit cell and in homogenous air (when a unit cell is removed). Bold points designate frequencies with the pressure distributions given in Table 1 and 2.

*Table 1. **"Fixed channel" case ("Fixed unit cell" case):** Pressure distributions around the 1st band gap for the labyrinthine metamaterial unit cells of the 3 iteration levels. Red and blue colors represent maximum and minimum pressure, while green color indicates (almost) zero pressure. The frequencies in brackets are referred to the "fixed unit cell" case.*

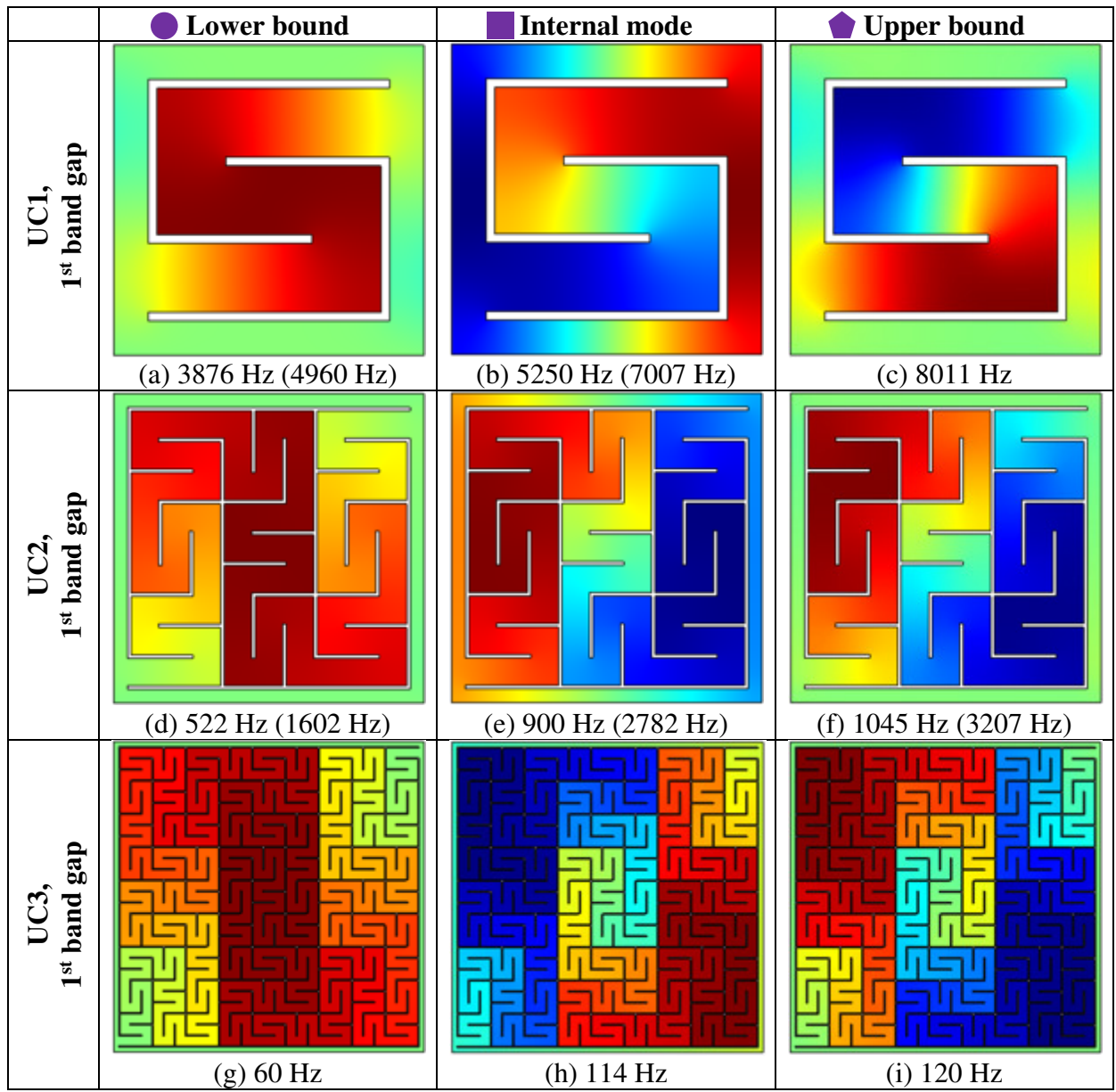

| | 🟣 **Lower bound** | 🟪 **Internal mode** | ⬠ **Upper bound** |
|---|---|---|---|
| **UC1, 1st band gap** | (a) 3876 Hz (4960 Hz) | (b) 5250 Hz (7007 Hz) | (c) 8011 Hz |
| **UC2, 1st band gap** | (d) 522 Hz (1602 Hz) | (e) 900 Hz (2782 Hz) | (f) 1045 Hz (3207 Hz) |
| **UC3, 1st band gap** | (g) 60 Hz | (h) 114 Hz | (i) 120 Hz |

*Table 2. "**Fixed channel**" case: Pressure distributions around the 2nd and 3rd band gaps for the labyrinthine metamaterial unit cells of the 2nd and 3rd iteration levels. Red and blue colors represent maximum and minimum pressure, and green color indicates (almost) zero pressure.*

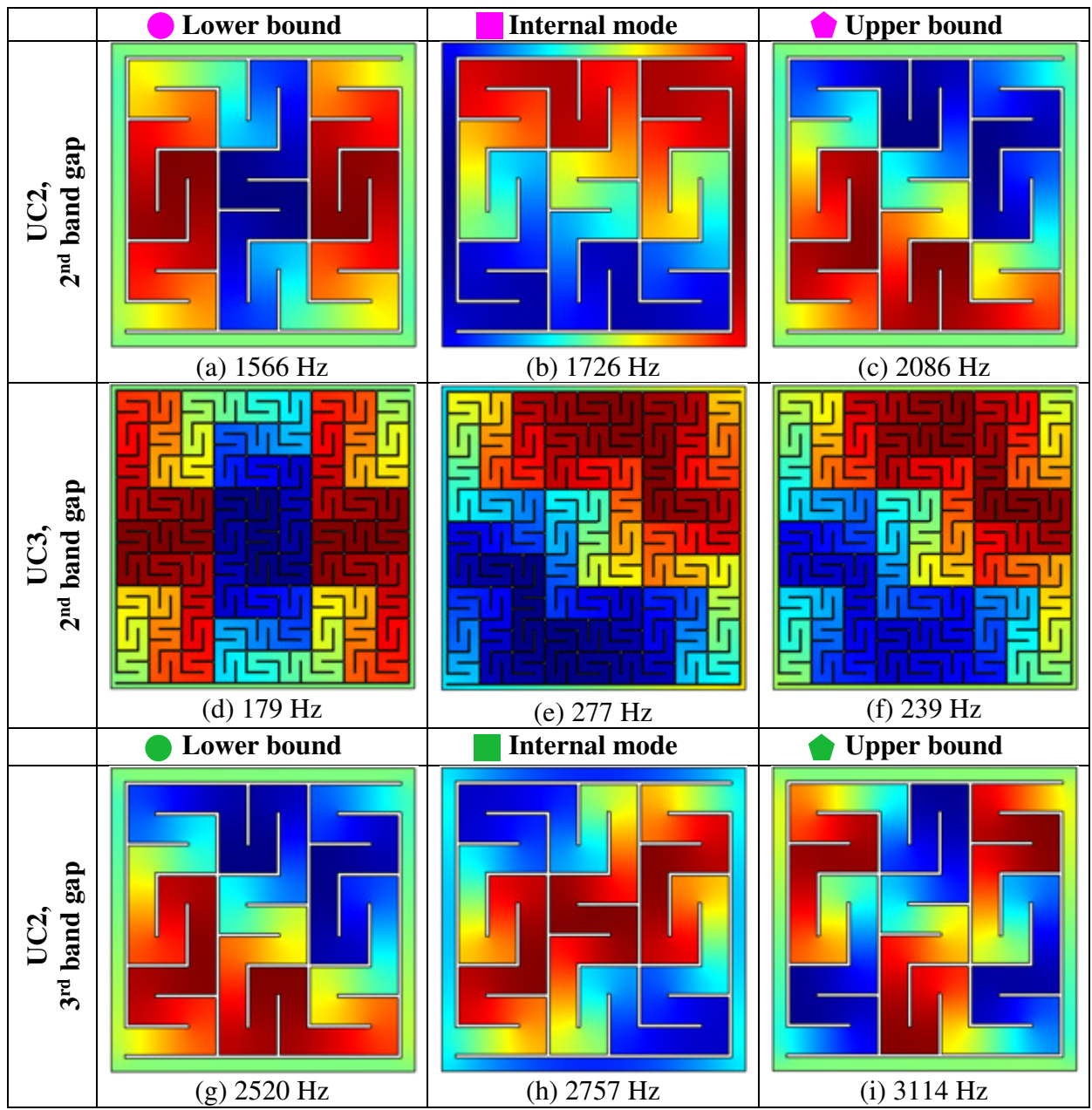

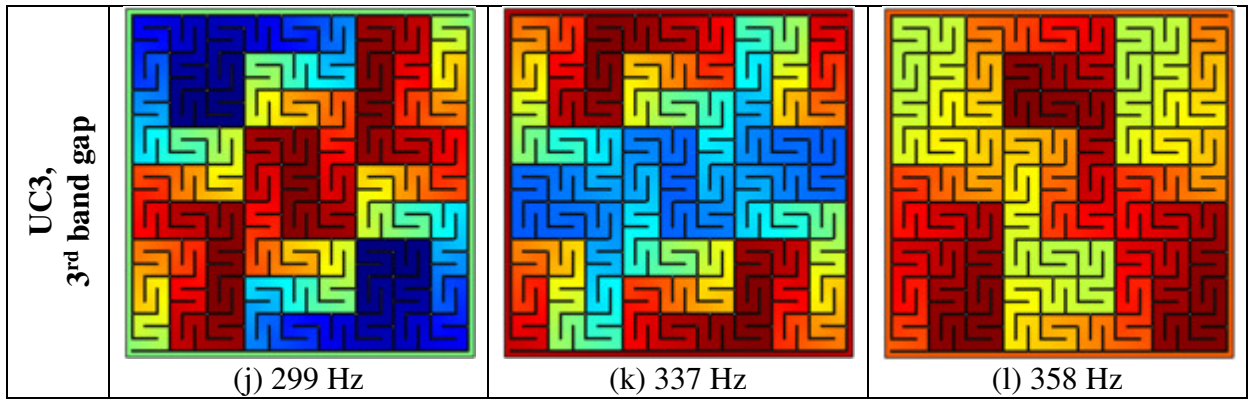

| UC3, 3rd band gap | (j) 299 Hz | (k) 337 Hz | (l) 358 Hz |

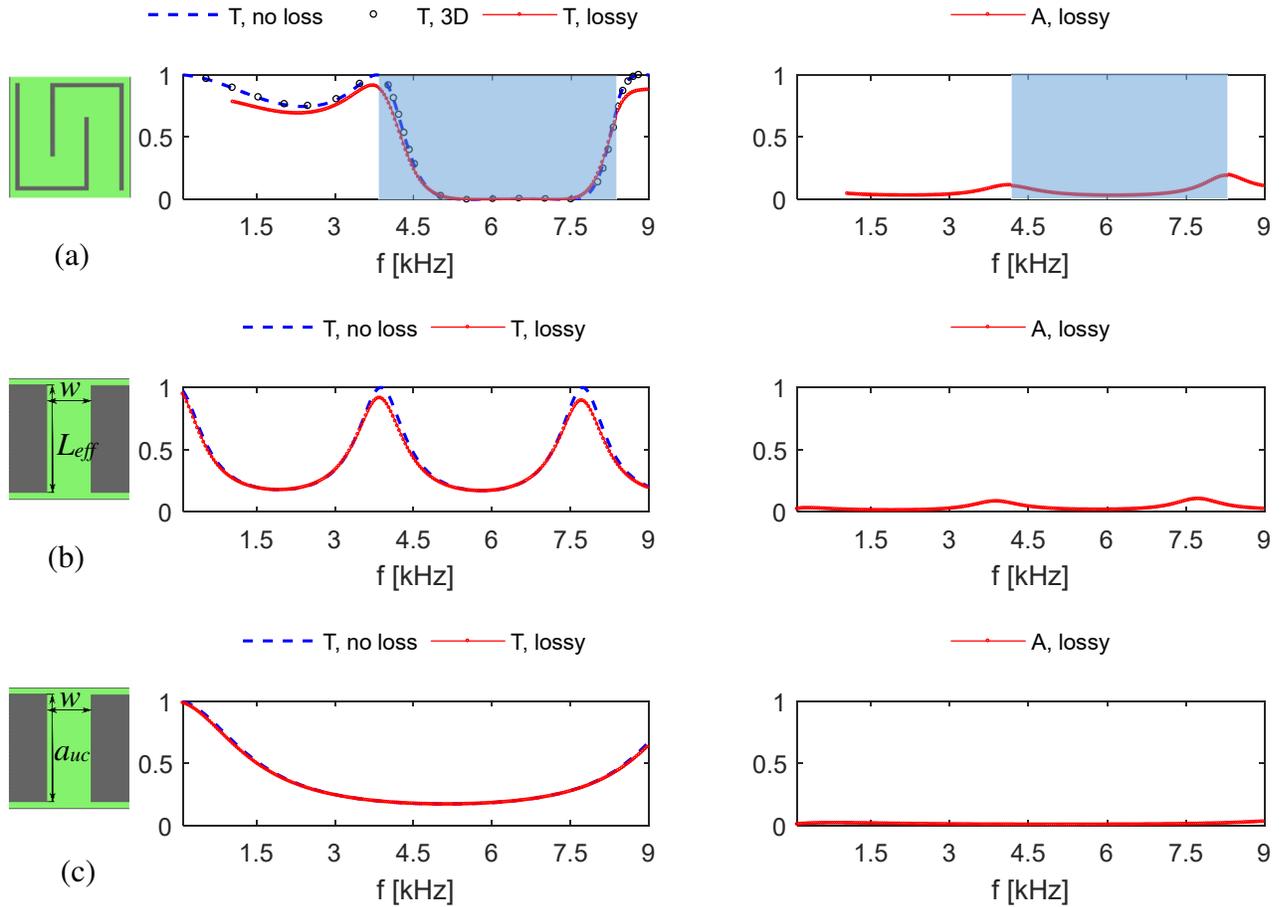

Figure 5. **"Fixed channel" case:** Transmission (T) and absorption (A) coefficients for acoustic waves in lossless (dashed line) and lossy (solid line) air through (a) a labyrinthine metamaterial UC1; (b) a straight slit of width $w$ = 4 mm and length $L_{eff}$ = 45.6 mm; (b) a straight slit of width $w$ = 4 mm and length $a_{uc}$ = 18 mm. Shaded regions indicate frequency a band gap shown in Fig. 4a. Circular markers in (a) indicate transmission coefficient values in lossless air for the corresponding 3D model of height $4a_{uc}$.

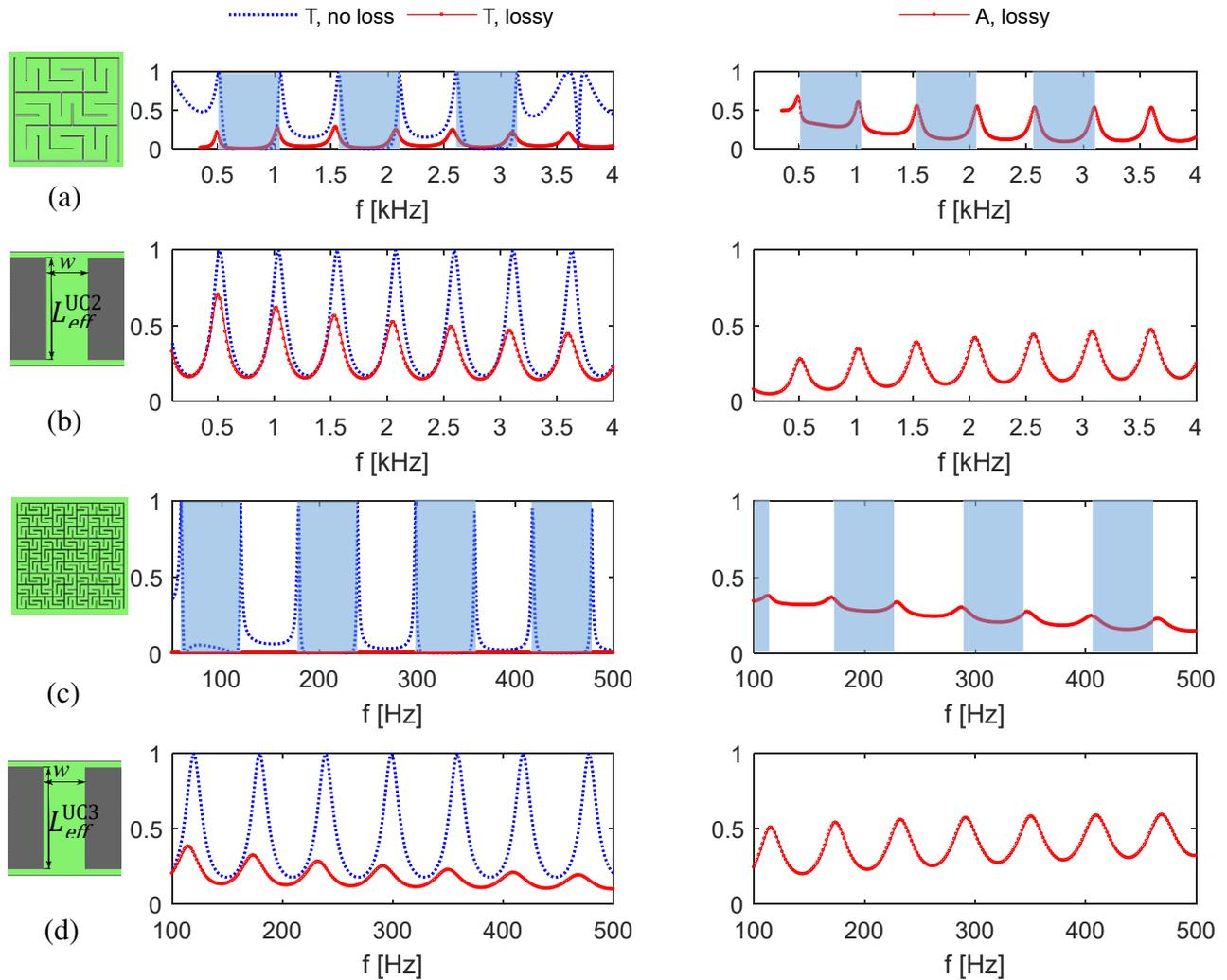

*Figure 6. "**Fixed channel**" case: Transmission (T) and absorption (A) coefficients for acoustic waves in lossless (dotted line) and lossy (solid line) air through (a) a labyrinthine metamaterial UC2 and (b) a straight slit of width w = 4 mm and length $L_{eff}$ = 328.5 mm; (c) a labyrinthine unit cell UC3 and (d) a straight slit of width w = 4 mm and length $L_{eff}$ = 2.871 m. Shaded regions indicate frequency band gaps shown in Fig. 4.*

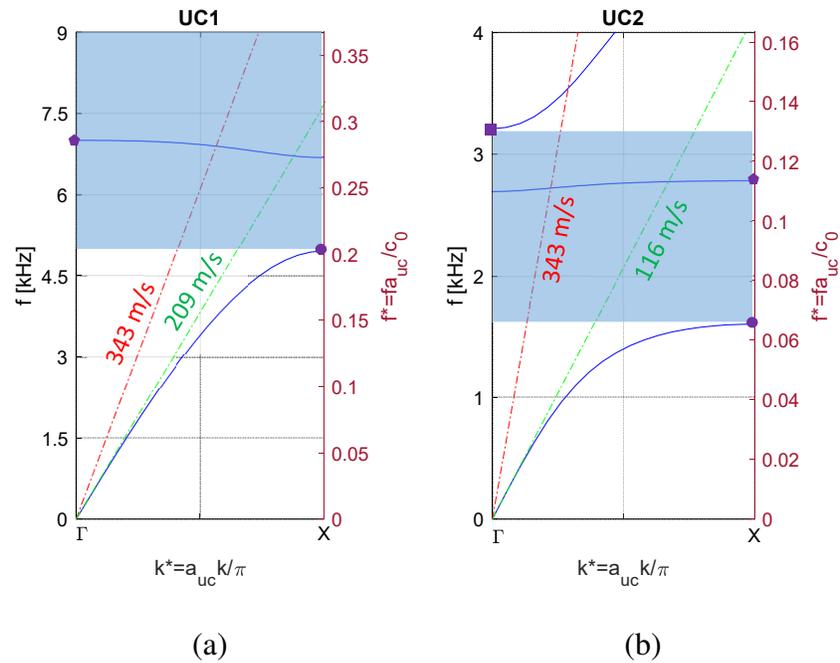

(a)                                    (b)

*Figure 7. **"Fixed unit cell" case**: Band structure diagrams for the unit cells UC1 and UC2 of fixed size a=14 mm with the channel width of 3 mm and 0.9 mm, respectively. Band gap frequencies are shaded. The slopes of the green and red dash-dot lines indicate the phase velocities of the fundamental pressure wave inside a unit cell and in homogeneous air (when a unit cell is removed).*

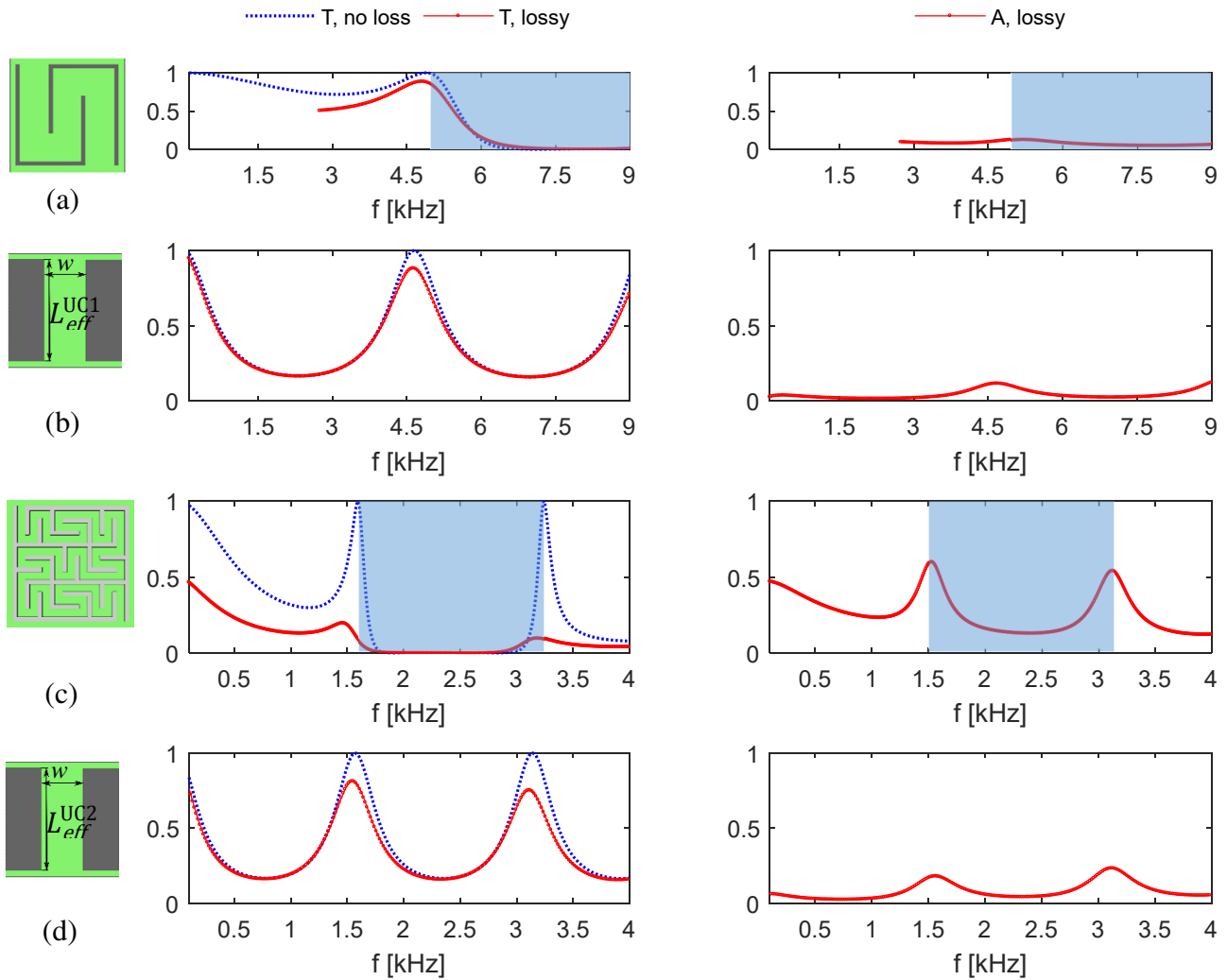

*Figure 8. **"Fixed unit cell"** case: Transmission (T) and absorption (A) coefficients for acoustic waves in lossless (dotted line) and lossy (solid line) air through (a) a labyrinthine unit cell UC1 and (b) a straight slit of width $w = 3$ mm and length $L_{eff} = 34.6$ mm; (c) a labyrinthine unit cell UC2 and (d) a straight slit of width $w = 0.9$ mm and length $L_{eff} = 107$ mm. Shaded regions indicate frequency band gaps shown in Fig. 7.*